\documentclass[twocolumn,aps,floatfix,showkeys,letterpaper,superscriptaddress]{revtex4}
\usepackage[utf8]{inputenc}
\usepackage[english]{babel}
\usepackage{graphicx}
\usepackage{hyperref}

\usepackage{braket}
\usepackage{siunitx}
\usepackage{xspace}
\newcommand{\BePlus}{$^9$Be$^+$\xspace}

\usepackage{float}

\usepackage{letltxmacro}
\LetLtxMacro{\ORIGselectlanguage}{\selectlanguage}
\makeatletter
\DeclareRobustCommand{\selectlanguage}[1]{%
	\@ifundefined{alias@\string#1}
	{\ORIGselectlanguage{#1}}
	{\begingroup\edef\x{\endgroup
			\noexpand\ORIGselectlanguage{\@nameuse{alias@#1}}}\x}%
}
\newcommand{\definelanguagealias}[2]{%
	\@namedef{alias@#1}{#2}%
}
\makeatother
\definelanguagealias{en}{english}
\definelanguagealias{En}{english}
\definelanguagealias{de}{german}
\definelanguagealias{EN}{english}

\begin{document}
\title{Optical stimulated-Raman sideband spectroscopy of a single \BePlus ion in a Penning trap}

\author{Juan~M.~Cornejo}
\email[]{cornejo-garcia@iqo.uni-hannover.de}
\affiliation{Institut für Quantenoptik, Leibniz Universität Hannover, Welfengarten 1, 30167 Hannover, Germany}
\author{Johannes~Brombacher}
\affiliation{Institut für Quantenoptik, Leibniz Universität Hannover, Welfengarten 1, 30167 Hannover, Germany}
\author{Julia~A.~Coenders}
\affiliation{Institut für Quantenoptik, Leibniz Universität Hannover, Welfengarten 1, 30167 Hannover, Germany}
\author{Moritz~von~Boehn}
\affiliation{Institut für Quantenoptik, Leibniz Universität Hannover, Welfengarten 1, 30167 Hannover, Germany}
\author{Teresa~Meiners}
\affiliation{Institut für Quantenoptik, Leibniz Universität Hannover, Welfengarten 1, 30167 Hannover, Germany}
\author{Malte~Niemann}
\affiliation{Institut für Quantenoptik, Leibniz Universität Hannover, Welfengarten 1, 30167 Hannover, Germany}
\author{Stefan Ulmer}
\affiliation{RIKEN, Ulmer Fundamental Symmetries Laboratory, 2-1 Hirosawa, Wako, Saitama 351-0198, Japan}
\affiliation{Institut für Experimentalphysik, Heinrich Heine Universität Düsseldorf, Universitätsstr. 1, 40225 Düsseldorf, Germany}
\author{Christian Ospelkaus}
\affiliation{Institut für Quantenoptik, Leibniz Universität Hannover, Welfengarten 1, 30167 Hannover, Germany}
\affiliation{Physikalisch-Technische Bundesanstalt, Bundesallee 100, 38116 Braunschweig, Germany}

\begin{abstract}
We demonstrate optical sideband spectroscopy of a single \BePlus ion in a cryogenic 5 Tesla Penning trap using two-photon stimulated-Raman transitions between the two Zeeman sublevels of the $1s^{2}2s$ ground state manifold. By applying two complementary coupling schemes, we accurately measure Raman resonances with and without contributions from motional sidebands. From the latter we obtain an axial sideband spectrum with an effective mode temperature of (3.1 $\pm$ 0.4)~mK. This results are a key step for quantum logic operations in Pennings traps, applicable to high precision matter-antimatter comparisons tests in the baryonic sector of the standard model.	
\end{abstract}
\keywords{Penning traps, laser cooling, stimulated Raman transitions, sideband spectroscopy}

\maketitle
Laser cooling of atomic ions in electromagnetic fields is a widely used and well-established technique~\cite{leibfried_quantum_2003}. It is application in radio-frequency (rf) traps is well known for optical clocks~\cite{ludlow_optical_2015}, quantum computation~\cite{kielpinski_architecture_2002} or quantum logic spectroscopy~\cite{schmidt_spectroscopy_2005}. However, it is less wide-spread in Penning traps because not many atomic species of interest are laser coolable; the ones which are suffer from the large Zeeman shifts caused by the high magnetic trapping fields of up to several Tesla~\cite{jordan_near_2019, gutierrez_trapsensor_2019}. In addition, the metastable magnetron motion \cite{brown_geonium_1986} leads to additional complications. Despite these challenges, there is growing interest in laser cooling for high-precision experiments in Penning traps because of the potential for improved accuracy and precision~\cite{cornejo_quantum_2012, cornejo_quantum_2021, baker_sympathetic_2021, will_sympathetic_2022}. For non laser-coolable ion species, such as protons, antiprotons or highly charged ions, sympathetic cooling techniques can be implemented, where the ion of interest is cooled by coupling it to laser-coolable ions~\cite{heinzen_quantum-limited_1990}. In addition, this could be used for motional ground state cooling, enabling the use of quantum logic protocols developed in rf traps~\cite{wolf_non-destructive_2016} for a wide range of applications in Penning traps experiments~\cite{wolf_motional_2019, wineland_experimental_1998-1, cornejo_quantum_2021}. High-precision matter-antimatter comparisons in the baryonic sector will strongly benefit from laser cooling~\cite{smorra_base_2015, bohman_sympathetic_2018, niemann_cryogenic_2019, will_sympathetic_2022}. Here, limitations due to cryogenic particle temperatures can be overcome by lower final temperatures and faster cooling times~\cite{bohman_sympathetic_2021}. Furthermore, if ground state cooling is achieved, quantum logic techniques for high-fidelity spin state detection could be implemented~\cite{cornejo_quantum_2021}.

Here, we demonstrate for the first time quantum spin-state manipulation of a single laser-cooled \BePlus ion in a Penning trap system using a two-photon stimulated-Raman process. In addition, the motional sideband spectrum of a single \BePlus ion is presented for the first time in a Penning trap. 

In a Penning trap, ions are confined by a superimposed quadrupole electric and axial magnetic field $\mathrm{\textbf{B}}$~\cite{brown_precision_1982}. The particle trajectory is a superposition of an axial motion along the magnetic field lines with characteristic frequency $\nu_z$ and two radial motions (magnetron and modified cyclotron) perpendicular to $\vec B$, with frequencies $\nu_-$ and $\nu_+$, respectively. The invariance theorem $\nu_{c}^2=\nu_+^2+\nu_z^2+\nu_-^2$~\cite{brown_precision_1982} links these frequencies to the free cyclotron frequency $\nu_{c} = qB/(2\pi m)$, which depends on $B$ and the charge-to-mass ratio $q/m$. The radial frequencies are given by $\nu_{\pm}=(\nu_{c}\pm \nu_{1})/2$, where  $\nu_{1}=\sqrt{\nu_{c}^2 - 2\nu_{z}^2}$, and the axial frequency by
\begin{equation}
	\label{z_freq}
	\nu_z=\frac{1}{2\pi}\sqrt{\frac{q\,V_{R}\,2C_2}{m}}.
\end{equation}
Here $V_{R}$ is the voltage applied to the trap electrodes and $C_2$ is a coefficient depending on trap geometry~\cite{gabrielse_cylindrical_1984}. 

\begin{figure}[t]
	\centering
	\includegraphics[width=1.0\columnwidth]{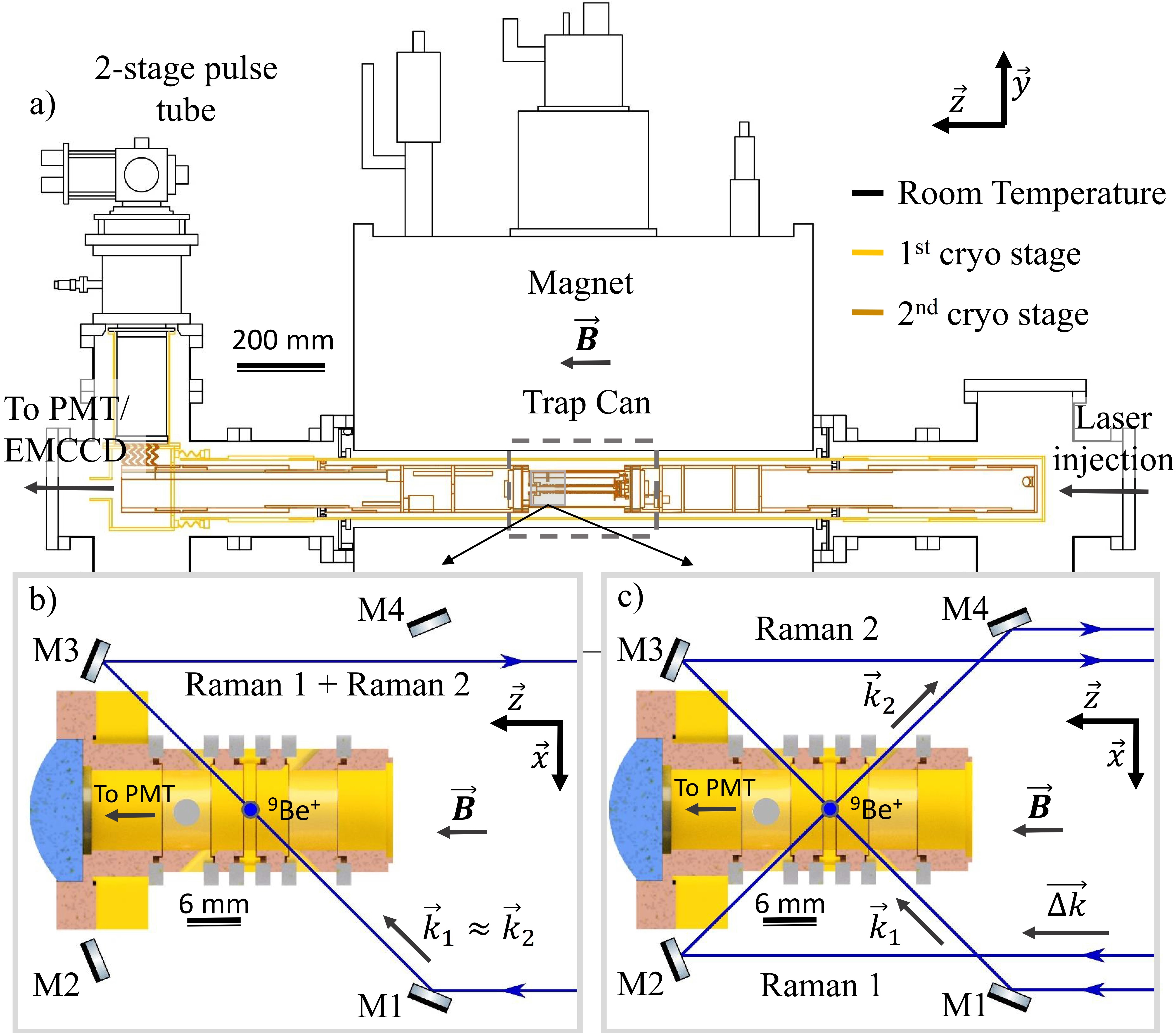}
	\caption{a) Cut-view of the main experimental setup, including the 5~T superconducting magnet as well as the ultra-low vibration cryocooler. The $1^{\mathrm{st}}$ and $2^{\mathrm{nd}}$ cryo stages have nominal working temperatures of 40~K and 4~K. The trap can, where the trap system is located, is at the center of the magnet. b) and c) show cut views of the Be trap in the co-propagating and crossed-beam laser configurations. Mirrors M1-M4 are used to guide laser beams into the trap.}
	\label{fig:setup}
\end{figure}

\begin{figure}[b]
	\centering
	\includegraphics[width=1.0\columnwidth]{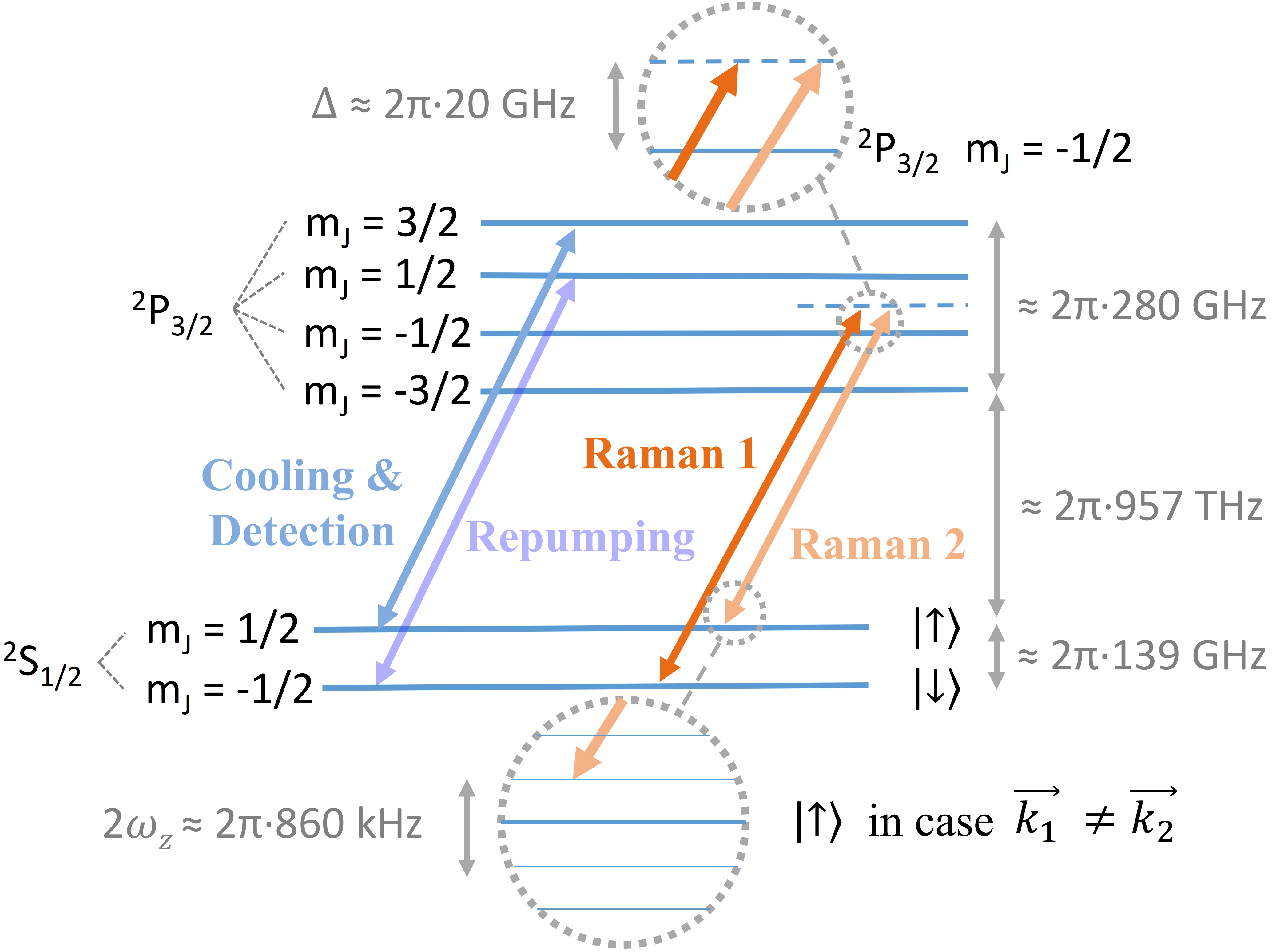}
	\caption{Simplified energy level scheme of \BePlus at $B=5\,\mathrm{T}$. Transitions for Doppler cooling and repumping are shown in dark and clear blue. Raman transitions are shown in dark and clear orange for Raman 1 and 2. The circle at the bottom indicates the different sideband transitions due to the interaction of the Raman lasers in crossed-beam configuration with the axial motion (energy levels not to scale).}
	\label{fig:energy}
\end{figure}

Doppler cooling of the axial and modified cyclotron modes in a Penning trap does not differ from the standard technique used in rf traps~\cite{leibfried_quantum_2003}. However, for the magnetron mode, understanding ``cooling'' as a reduction of the motional amplitude, it is necessary to apply either a radial intensity gradient~\cite{itano_laser_1982} or axialization~\cite{phillips_dynamics_2008}. Here, the former is used. After Doppler cooling, the ion is prepared in a thermal state with an occupation probability of a Fock state $\ket{n}$ given by $P(n) = \bar{n}^n / (\bar{n}+1)^{n+1}$~\cite{mavadia_optical_2014}. The mean phonon number $\bar{n}$ can refer to any of the modes of the trapped ion, in particular for the axial mode $\bar{n}_z \approx k_B T_z/h\nu_z$, where $h$ is Planck's constant, $k_B$ the Boltzmann constant and $T_z$ the axial temperature of the ion. The latter will be close to the Doppler limit of $\approx0.5\,\mathrm{mK}$ for a single \BePlus ion.

In order to achieve temperatures below the Doppler limit, sideband laser cooling can be performed~\cite{diedrich_laser_1989}. To implement this, it is necessary to couple the motional and internal states of the trapped ions. This can be performed using two-photon stimulated-Raman transitions through a third optical level~\cite{wineland_experimental_1998-1, paschke_versatile_2019, mielke_139_2021}. In this case, the interaction is restricted to a resonance condition which follows from the effective Hamiltonian $H \propto \mbox{exp}\{i[(\vec{k}_1-\vec{k}_2)\cdot\vec{r}-(\nu_1 - \nu_2)t]\}$~\cite{wineland_experimental_1998-1}, where $\vec{k}_1$, $\vec{k}_2$ and $\nu_1$, $\nu_2$ are the wavevectors and frequencies of the Raman laser beams, respectively. The resonance condition $(\nu_1 - \nu_2) - \nu_0 = 2\pi\nu_z(n_1 - n_2)$, where $\nu_0$ is the transition frequency, results in a motional sideband spectrum with a weighted average of the coupling strengths of the corresponding Fock states $\ket{n_1}$ and $\ket{n_2}$~\cite{wineland_experimental_1998-1}. The case $n_1 = n_2$ corresponds to the carrier transition, and the cases $n_1 > n_2$ and $n_1 < n_2$ correspond to blue and red sideband transitions, respectively. For a mean phonon number much larger than 1, the averaged spectra will follow a Doppler broadened envelope as in the classical picture~\cite{wineland_laser_1979}, which can be used to determine the particle temperature~\cite{mavadia_optical_2014, mielke_139_2021}. 

Our cryo-mechanical setup is illustrated in Fig.~\ref{fig:setup}. The Penning trap is located in the center of a superconducting magnet with $B=5\,\mathrm{T}$. The trap system is cooled down to $\approx$\,6\,K using a two-stage cryocooler. Since the trap can is also vacuum sealed, a pressure lower than $10^{-14}$~mbar is estimated. More details can be found in \cite{niemann_cryogenic_2019}. All laser beams are injected from upstream (the right side in Fig.~\ref{fig:setup}), while the downstream (the left side in Fig.~\ref{fig:setup}) is used to monitor ion fluorescence using a photomultiplier tube or an electron-multiplying charge-coupled device camera. 

Figure~\ref{fig:energy} shows the $^2P_{3/2}$ and $^2S_{1/2}$ sublevels of a single \BePlus ion at $B=5\,\mathrm{T}$ ($^2P_{1/2}$ not shown for simplicity). For Doppler cooling and fluorescence detection, the $\ket{^2S_{1/2}, m_j=+1/2}$ $\rightarrow$ $\ket{^2P_{3/2}, m_j=+3/2}$ transition is used. A repump laser resonant with the $\ket{^2S_{1/2}, m_j=-1/2}$ $\rightarrow$ $\ket{^2P_{3/2}, m_j=+1/2}$ transition depletes the $\ket{^2S_{1/2}, m_j=-1/2}$ state and initializes the ion in $\ket{^2S_{1/2}, m_j=+1/2}$. Via two-photon induced Raman transitions (see Fig.~\ref{fig:energy}), coherent state transfer between the $^2S_{1/2}$ sublevels can be achieved. A detuning of $\approx20\,\mathrm{GHz}$ from the $\ket{^2S_{1/2}, m_j=-1/2}\equiv\ket{\downarrow}$ $\rightarrow$ $\ket{^2P_{3/2}, m_j=-1/2}$ transition is used for Raman 1, and from $\ket{^2S_{1/2}, m_j=+1/2}\equiv\ket{\uparrow}$ $\rightarrow$ $\ket{^2P_{3/2}, m_j=-1/2}$ for Raman 2. All four ultraviolet (UV) laser beams are generated using infrared (IR) fiber lasers, sum frequency generation (SFG) and a second harmonic generation~\cite{hannig_highly_2018,wilson_750-mw_2011}. The frequency difference of the Raman lasers is matched with the Zeeman splitting of $139\,\mathrm{GHz}$ ($B=5\,\mathrm{T}$) of the \BePlus $S_{1/2}$ sublevels by phase locking two IR lasers using an electro-optic modulator modulated~\cite{mielke_139_2021}.

Although our trap system consists of several traps~\cite{niemann_cryogenic_2019, cornejo_quantum_2021}, we will now focus on the so-called ``Be trap'', a five-electrode compensated and orthogonal cylindrical trap of $9\,\mathrm{mm}$ inner diameter shown in Fig.~\ref{fig:setup}. Holes in the electrodes and several mirrors (M1, M2, M3 and M4) allow the injection of laser beams under an angle of $45^\circ$ with respect to $\vec B$. The Doppler laser as well as the repump laser beams are injected using mirrors M1 and M3. For the so-called ``co-propagating'' configuration (see Fig.~\ref{fig:setup}.b)), both Raman beams are applied using mirrors M1 and M3. Because the wavevectors of both beams are aligned with similar magnitude, the interaction with the axial motion can be neglected. However, for the ``crossed-beam'' configuration of Fig.~\ref{fig:setup}.c), Raman 1 now uses mirrors M2 and M4, resulting in a wavevector difference $\vec{\Delta k} = \vec{k_1}-\vec{k_2}$ along $\vec B$. This allows an interaction with the internal and vibrational states.

Ions are loaded using a 532~nm ablation laser beam impinging on a beryllium target (gray circle of figures~\ref{fig:setup}b) and ~\ref{fig:setup}c)) embedded into a trap electrode. A single ion is prepared by selective splitting and ejection using the electric trapping potential~\cite{niemann_cryogenic_2019, meiners_transport_2022}. The ion is cooled using a Doppler beam with a power stabilized to $400\,\mu$W, a beam waist of $\approx150\,\mu$m and a vertical offset of $\approx100\,\mu$m from the trap center. During cooling, the repump laser (power stabilized to $40\,\mu$W and beam waist $\approx150\,\mu$m) is used to avoid decay into a dark state, which can happen as a result of the cooling laser off-resonantly driving a transition to e.g.~$\ket{^2P_{3/2}, m_j=+1/2}$ with subsequent decay to $\ket{^2S_{1/2}, m_j=-1/2}$ (note that because of its angle relative to the magnetic field, the cooling laser can in principle induce transitions other than $\Delta m_j=+1$, even for perfectly circular polarization). An axial trap frequency of $435.7\,\mathrm{kHz}$ is achieved by applying $-20\,\mathrm{V}$ to the ring electrode and $-17.6\,\mathrm{V}$ to the correction electrodes~\cite{mielke_thermometry_2021}. After turning the cooling lasers off, the Raman transition is probed by turning on the Raman lasers for the interaction time. Afterwards, the Doppler laser is pulsed on for spin state detection. In case of the $\ket{\uparrow}$ state, fluorescence will be observed, and no photons otherwise. Finally, the repump laser is used together with the cooling laser to re-initialize the ion in $\ket{\uparrow}$, and to start the cooling process again.

\begin{figure}[b]
	\centering
	\includegraphics[width=1.0\columnwidth]{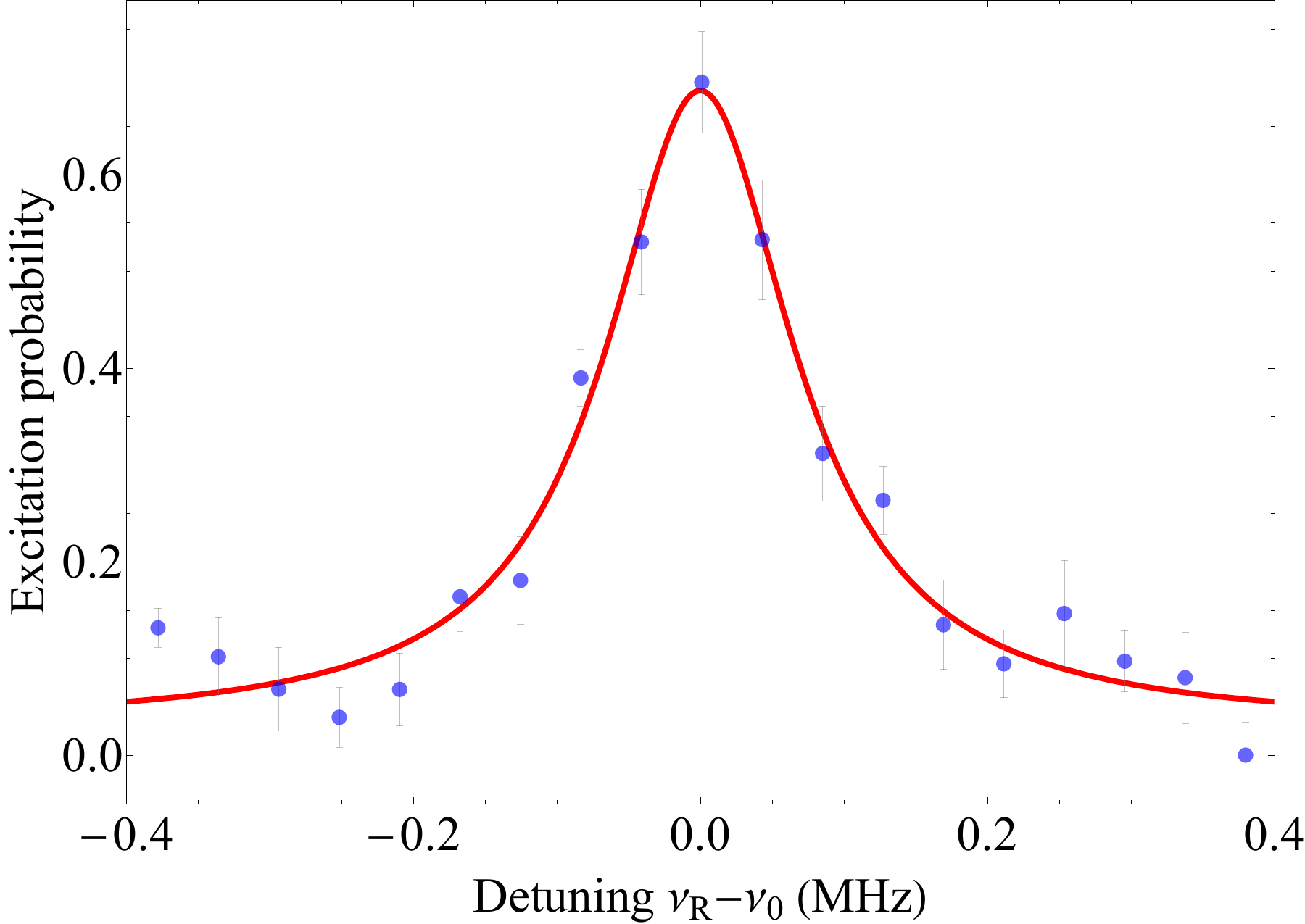}
	\caption{Blue dots: $\ket{^2S_{1/2}, m_j=+1/2}$ $\rightarrow$ $\ket{^2S_{1/2}, m_j=-1/2}$ transition probability as a function of Raman laser frequency detuning in the co-propagating configuration. The red line is a Lorentzian fit to the data. Each data point is obtained from 180 individual measurements.}
	\label{fig:resonance}
\end{figure}

For these single-ion measurements, it was necessary to tightly control the ion motions using Doppler cooling. ``Jumps'' in the fluorescence counts of individual ions or small ion clouds were observed earlier in the setup~\cite{niemann_cryogenic_2019}. Using only single ions to avoid changes in the cloud conformation, improvements of the cleaning process for loading and a more accurate measurement of the repump transition~\cite{mielke_thermometry_2021} improved this somewhat. Still, \BePlus has the smallest mass among readily laser coolable ions with comparable cooling transition natural linewidth. This results in a larger Doppler broadening for a given temperature and emphasizes higher-order effects in the cooling dynamics, in particular for smaller detunings. Linear approximations in the cooling dynamics become more valid with larger Doppler laser detuning~\cite{thompson_simple_2000}. The non-linear behavior is also less an issue for ion crystals because the strong Coulomb interaction in the ion crystal, which suppresses free-ion mobility. Also, the overall radial orbit of a multi-ion radial crystal is larger, favoring magnetron ``cooling'' by the radial intensity gradient. This allowed us to work with the nominally optimal Doppler cooling detuning of $-10\,\mathrm{MHz}$ for ion clouds in~\cite{mielke_139_2021}. A key step forward for the present single-ion experiments was a larger Doppler laser detuning of $-20\,\mathrm {MHz}$ which, at the price of a slightly higher nominal temperature, still allows a much more reproducible motional state preparation. Any remaining cases of motion instability, where the signal vanishes for any detuning of the Raman lasers, are discarded based on a threshold value in the number of counts per scan. In our present setup, we do not accept rejection rates larger than 10\%. Above this threshold, we re-align the laser beams. 

\begin{figure}[t]
	\centering
	\includegraphics[width=1.0\columnwidth]{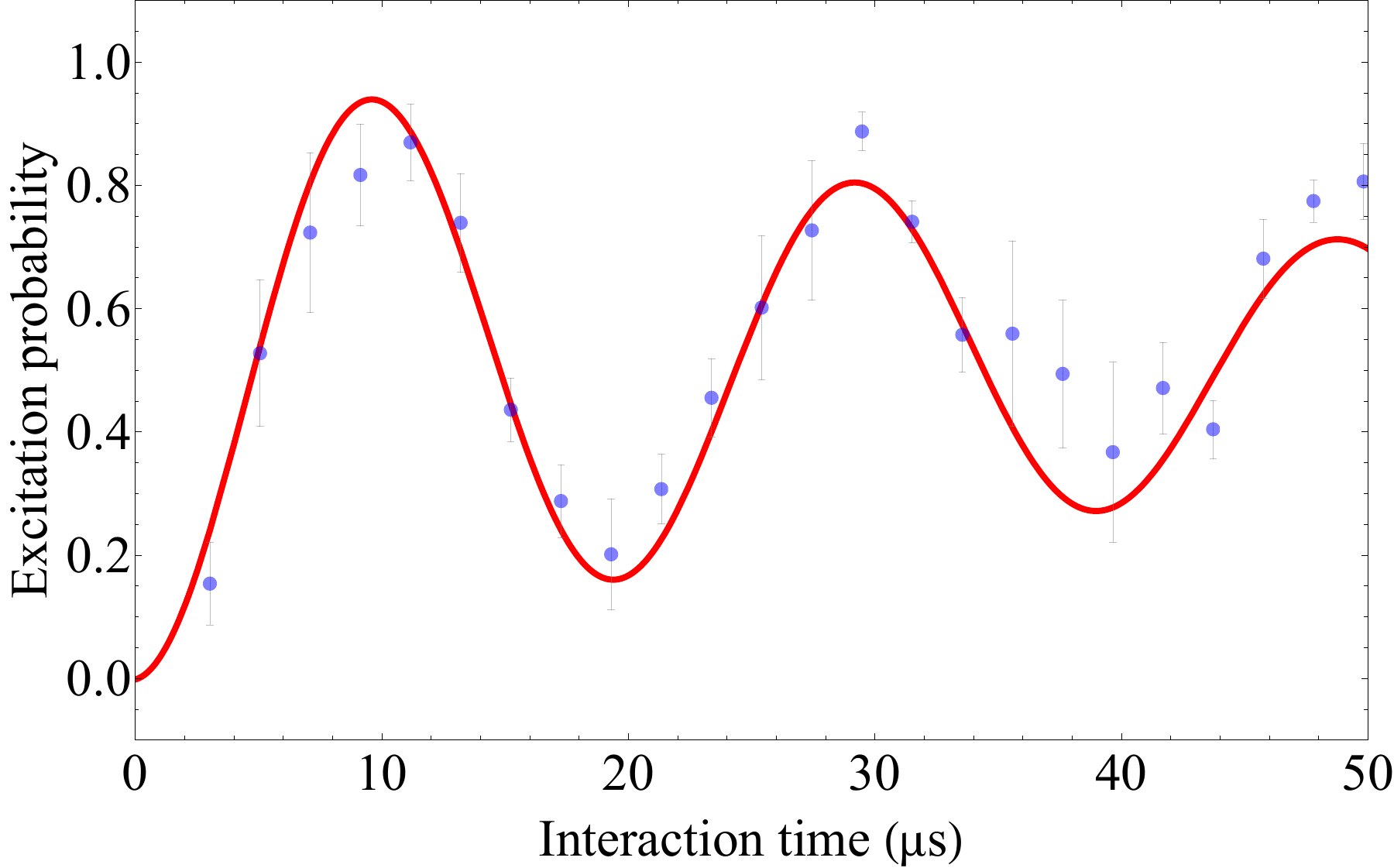}
	\caption{Excitation probability for the $\ket{^2S_{1/2}, m_j=+1/2}$ $\rightarrow$ $\ket{^2S_{1/2}, m_j=-1/2}$ transition in the co-propagating laser beam configuration as a function of interaction time (30 measurements per data point). The red line is a fit to the data based on an exponentially decaying sinusoid.}
	\label{fig:rabi}
\end{figure}

\begin{figure*}[t]
	\centering
	\includegraphics[width=2.0\columnwidth]{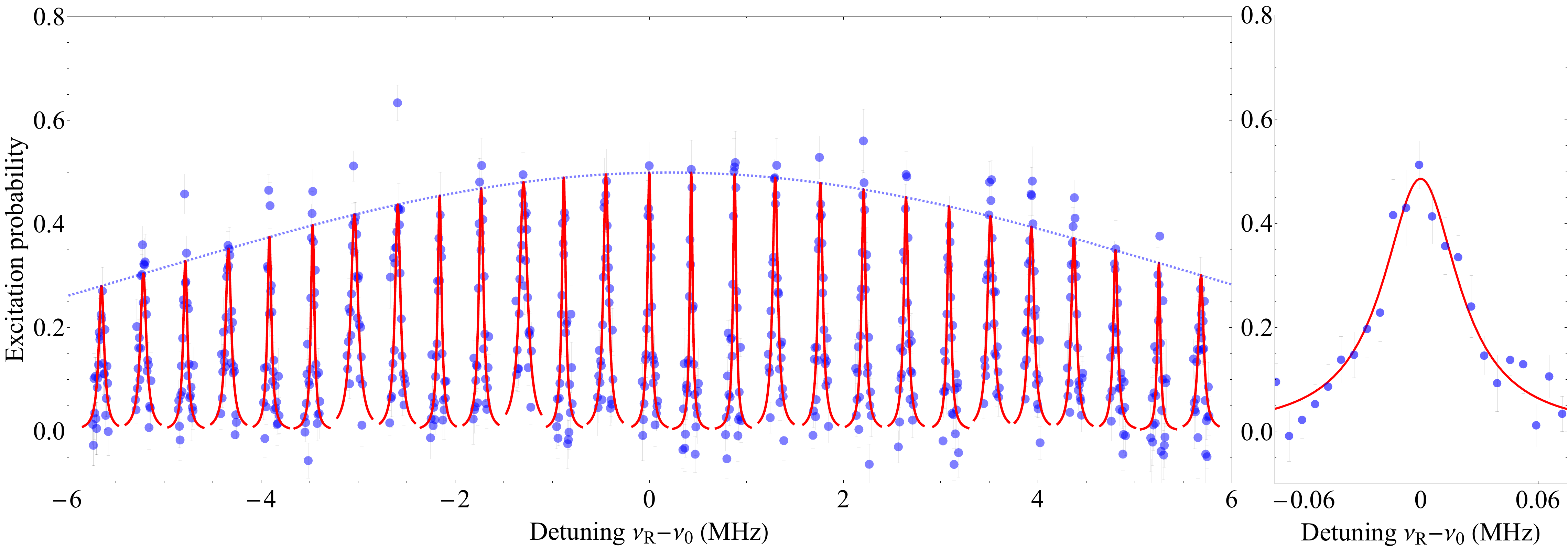}
	\caption{Left panel: Resolved-sideband spectrum (transition probability as a function of Raman laser detuning) observed with the crossed-beam configuration. Each sideband was scanned individually, with 100 measurements per data point. Red line: Lorentzian functions with Gaussian envelope fitted to the data. Right panel: Carrier transition probability to illustrate the fit of an individual line. }
	\label{fig:spectra}
\end{figure*}

Figure~\ref{fig:resonance} shows a scan of the Raman laser frequency difference relative to the expected spin-flip resonance for the co-propagating configuration. The ion is expected to undergo transitions from $\ket{\uparrow}$ to $\ket{\downarrow}$, with negligible interaction with the axial motional mode. $\nu_{R}$ is the Raman lasers' frequency difference, and $\nu_{0}=138912.278\,\mathrm{MHz}$. Here, the ion was laser-cooled for 5~ms and a Raman interaction time of $75\,\mu\mathrm{s}$ was used. Raman laser powers were stabilized to 7.5\,mW and 2\,mW for Raman lasers 1 and 2, respectively. The difference in power is due to the required polarization of each Raman transition~\cite{mielke_139_2021}. For a total of 18 scans, each data point was measured 10 times. The excitation probability is obtained as one minus the normalized number of photon counts.

Coherent state manipulation is demonstrated by driving Rabi oscillations (see Fig.~\ref{fig:rabi}). Here, the frequency difference between the Raman lasers was fixed to $\nu_0$, and the interaction time was scanned. The fit yields a frequency of (49.6 $\pm$ 3.0)~kHz and a decay time of (52 $\pm$ 14)~$\mu$s. Small variations in position of the Raman lasers likely cause the decay, as the the resulting changes of laser intensity at the ion affect the Rabi rate and AC Stark shifts~\cite{ludlow_optical_2015}. 

In the crossed-beam configuration, the wavevector difference of the Raman lasers along $\vec B$ allows observation of the sideband spectrum by coupling the internal degrees of freedom to the axial vibrational mode. The left panel of Fig.~\ref{fig:spectra} shows the sideband spectrum of a single \BePlus ion. The carrier transition and thirteen blue and red sidebands were measured individually. For each sideband, a total of 10 scans were performed at 10 experiments per data point. Here, $\nu_{0}=138913.014\,\mathrm{MHz}$ due to a magnetic-field drift over several weeks between Figs.~\ref{fig:resonance} and~\ref{fig:spectra}. The ion was laser-cooled for $1\,\mathrm{ms}$ and an interaction time of $700\,\mu\mathrm{s}$ was used (laser powers $\approx 3\,\mathrm{mW}$ and $1\,\mathrm{mW}$ in Raman 1 and 2, respectively). 

The Gaussian envelope of Fig.~\ref{fig:spectra} has an FWHM of $\nu_D=(12.8\pm0.8)\,\mathrm{MHz}$. Using $T_z=m\lambda^2\nu_D^2/(8~ln2~k_B)$ for a Doppler-broadened spectrum~\cite{mielke_thermometry_2021}, where $\lambda$ is the laser wavelength, an axial temperature after cooling of $T_z=(3.1 \pm 0.4)\,\mathrm{mK}$ is determined. This value is a factor of $\approx 6$ larger than expected from the Doppler limit, and a factor of $\approx 2$ compared to~\cite{mielke_139_2021} measured with an ion cloud. We estimate that a factor of $\approx 1.5$ is due to the detuning of $-20\,\mathrm{MHz}$ instead of $-10\,\mathrm{MHz}$ as in~\cite{mielke_139_2021}. The remaining factor of 4 is likely due to remaining motional instabilities and cooling dynamics nonlinearities. Introduction of a pure axial laser beam combined with the axialization technique~\cite{phillips_dynamics_2008} may improve this further. The former will allow us to apply independent Doppler-laser detunings for the radial and axial motions, obtaining more precise control of the ion motions during cooling. For the latter, although we are able to couple the ion modes in the trap by means of a quadrupole fields, we are not able to observe a reduction in the final temperature of the ion, possibly due to poor overlap of the electrostatic and rf excitation fields~\cite{goodwin_sideband_2015}, an issue that can be fixed with apparatus upgrades allowing fine tuning of the rf excitation field. Nevertheless, the current status of the system allows us to reach a mean axial phonon number of $\bar{n}_z \approx 150 \pm 19$ with a single laser-cooled \BePlus ion. It will therefore be possible to implement resolved-sideband cooling to reduce the mean phonon number~\cite{goodwin_resolved-sideband_2016}.

In conclusion, we have demonstrated two-photon induced Raman transitions on a single \BePlus in a $5\,\mathrm{T}$ cryogenic Penning trap by phase-locking two UV-laser sources at an offset frequency of $\approx 139\,\mathrm{GHz}$. Transition frequency and Rabi rate have been determined with a co-propagating Raman laser beam configuration, and finally, a spectrum with 26 sidebands has been measured using a crossed-beam configuration, yielding an axial temperature of (3.1 $\pm$ 0.4)~mK. The temperature is several orders of magnitude below the cryogenic temperature of particles used in current experiments on matter-antimatter comparison tests~\cite{schneider_double-trap_2017}. Improvements on our experimental system will allow us to approach the axial Doppler limit closer and perform sideband cooling to ultimately reach the ground state of the axial mode, enabling the application of quantum logic techniques in Penning trap-based precision experiments.  

\begin{acknowledgments}
We are grateful for discussions with  J.J.\ Bollinger, R.C.\ Thompson and P.O.\ Schmidt. This work was supported by PTB, LUH, and DFG through the clusters of excellence QUEST and QuantumFrontiers as well as through the Collaborative Research Center SFB1227 (DQ-mat Project-ID 274200144) and ERC StG ``QLEDS''. We also acknowledge financial support from the RIKEN Pioneering Project Funding and the MPG/RIKEN/PTB Center for Time, Constants and Fundamental Symmetries.
\end{acknowledgments}

\bibliography{QLEDS_sideband_spectroscopy}

\end{document}